\documentclass[fleqn,usenatbib]{mnras}

\usepackage[T1]{fontenc}
\usepackage{graphicx}	
\usepackage{amsmath}	
\usepackage{amssymb}	

\usepackage{natbib} 
\usepackage{lineno,hyperref}
\usepackage{float}
\usepackage{xcolor}
\usepackage{booktabs}

\DeclareRobustCommand{\VAN}[3]{#2}
\let\VANthebibliography\thebibliography
\def\thebibliography{\DeclareRobustCommand{\VAN}[3]{##3}\VANthebibliography}

\newcommand\aoflagger{{SDP Flagger}} 
\newcommand\sdp{{SDP Flagger}} 
\newcommand\kat{KAT-7}
\newcommand\unet{U-Net}
\newcommand\rnet{R-Net}

\title[DeepRFI]{Deep Learning improves identification of Radio Frequency Interference}

\author[Vafaei Sadr et al.]{
Alireza Vafaei Sadr$^{1,2}$,
Bruce  A. Bassett$^{3,4,5,6}$, Nadeem Oozeer$^{3,5}$
\newauthor 
\;
Yabebal Fantaye$^{5}$ 
and Chris Finlay$^{3,4,5}$
\\~\\
$^{1}$ Institute for Research in Fundamental Sciences (IPM), P. O. Box 19395-5531, Tehran, Iran\\
$^2$ D\'epartement de Physique Th\'eorique and Center for Astroparticle Physics, University of Geneva\\
$^{3}$ South African Radio Astronomy Observatory (SARAO), 2 Fir Street, Observatory, Cape Town, 7925, South Africa\\
$^{4}$ Department of Maths and Applied Maths, University of Cape Town, Cape Town, South Africa\\
$^{5}$ African Institute for Mathematical Sciences, 6 Melrose Road, Muizenberg, 7945, South Africa\\
$^{6}$ South African Astronomical Observatory, Observatory, Cape Town, 7925, South Africa\\ 
}

\date{Accepted XXX. Received YYY; in original form ZZZ}

\pubyear{2020}

\begin{document}
\label{firstpage}
\pagerange{\pageref{firstpage}--\pageref{lastpage}}
\maketitle

\begin{abstract}
Flagging of Radio Frequency Interference (RFI) in time-frequency visibility data is an increasingly important challenge in radio astronomy. We present \rnet, a deep  convolutional ResNet architecture that significantly outperforms existing algorithms --  including the default MeerKAT RFI flagger, and  deep \unet\ architectures --  across all metrics including AUC, F1-score and MCC. We demonstrate the robustness of this improvement on both single dish and interferometric  simulations and, using transfer learning, on real data. Our \rnet\ model's precision is approximately  $90\%$ better than the current MeerKAT flagger  at $80\%$ recall and has a 35\% higher F1-score with no additional performance cost. 
We further highlight the effectiveness of transfer learning from a model initially trained on simulated MeerKAT data and fine-tuned on real, human-flagged, \kat\ data. Despite the wide differences in the nature of the two telescope arrays the model achieves an AUC of 0.91, while the best model without transfer learning only reaches an AUC of 0.67. We consider the use of phase information in our models but find that without calibration the phase adds almost no extra information relative to amplitude data only. 
Our results strongly suggest that deep learning on simulations, boosted by transfer learning on real data, will likely play a key role in the future of RFI flagging of radio astronomy data. 
\end{abstract}




\section{Introduction}

Radio astronomy observatories are driven to the quietest regions on earth in an attempt to escape the relentless contamination from man-made radio emission including satellites, television, radio, cell phones and aircraft. Despite this, contamination from Radio Frequency Interference (RFI) is often orders of magnitude stronger than the astronomical signals of interest and hence must be carefully removed from data. RFI will increasingly be a limiting factor for high quality science observations with current and planned radio telescopes, such as the MeerKAT, the South African precursor of the Square Kilometer Array (SKA) \citep{ellingson2005rfi}. 

Despite international and local regulations
to limit RFI contamination in areas where radio telescopes operate,
or adopting observational strategies to avoid known RFI sources,
RFI cleaning, flagging or masking procedures are needed to obtain
robust science results. Exploring the best way to remove post-correlation data significantly contaminated by RFI is the subject of this paper. 

Excising RFI is complicated by the fact that RFI can take an exceptionally wide range of forms, with diverse effects on different science goals such as mapping the HI spectral line as a function of redshift, continuum emission surveys and detection of new transients. The latter is particularly affected by RFI since it can appear on a wide range of characteristic timescales from nanoseconds to hours; for a review see \cite{fridman2001rfi}.

There have been numerous techniques proposed to address the excision or mitigation of RFI from observed data. These include Singular Vector Decomposition (SVD) \citep{offringa2010post}, Principle Component Analysis
\citep{zhao2013windsat}; {\sc cumsum} \citep{baan2004radio}, {\sc SumThreshold}
\citep{raza2002spatial,offringa2010post}, methods exploiting polarisation information \citep{yatawatta2020polarization} and  
finally  supervised machine Learning where the algorithm learns from classified examples of RFI; e.g. \cite{wolfaardt2016machine,mosiane2016radio}.

Machine leaninrg techniques have been proposed to address different problems in astronomy \citep{harp2019machine,vafaei2019deepsource,nieto2019ctlearn,connor2018applying}. This paper presents a method in the machine learning category, exploiting the powerful ability of deep neural networks – and convolutional neural networks (CNN) in particular – to learn the relevant features used to do the classification.

In the context of RFI flagging, 
deep learning has been used on simulated single dish data \citep{akeret2017radio} where a \unet\ architecture delivered  better results than traditional methods. Our paper extends this result by exploring new deep architectures and, in particular, now demonstrates superiority also on  simulated interferometric data sets. Other studies have also looked to deep learning to help with RFI \citep{burd2018detecting,czech2018cnn,sclocco2020real}, including in the context of modern facilities such as FAST \citep{yang2020deep} and HERA \citep{kerrigan2019optimizing}, including the use of generative models \citep{GAN_RFI}.

There are two big challenges to using supervised deep learning for RFI: the first is that we need lots of data with labels (RFI or non-RFI). It is extremely tedious for humans to accurately label many different baselines (in the case of MeerKAT over 2000 baselines when the full array of 64 dishes is being used). In addition, different humans will disagree on the boundary of the RFI and hence we do not have access to the ground truth in the data.

On the other hand, a simulator for RFI for radio interferometers was not available, so we had to create it. This solves both the problem of the ground truth (we know what signals were put into the data) and the volume of data (we can simulate arbitrarily large amounts). However, it leads to another problem: how do we know if our wonderful model trained on simulated data, is any good on real data? Further, if it turns out to not be very good, how do we improve the model? We solve this problem by using transfer learning \citep{tan2018survey} onto real KAT-7 data which was human labeled. We find excellent results that outperform both the U-Net and traditional approaches.

The paper is organized as follows: in Sec.~\ref{data} we describe the
simulation and data sets we use; in Sec.~\ref{model} we describe 
the applied RFI flagging algorithms and explain the details of the different CNN
architectures we considered; in Sec.~\ref{results} we give the results and finally conclude in Sec.~\ref{conclusion}.


\section{Data sets}
\label{data}

In order to compare the performance of the convolutional neural network based architecture against the traditional RFI flaggers we use three data sets in our analysis: two simulated sets and one based on real astronomical observations. 

\subsection{Single-Dish  simulations}

The first and simplest data set is 13-months of simulated single-dish data  using the HIDE \& SEEK package \citep{akeret2017hide}  matching the characteristics of the Bleien Observatory 276 frequency channels with 1 MHz bandwidth between 990 MHz and 1260 MHz. We used similar experimental setup and RFI simulation parameters to those in \cite{akeret2017radio}.  The training, validation and test sets consisted of 11 months, 1 month and 1 month respectively. 
A single day of simulated HIDE data can be seen in the top panel of in Fig. \ref{fig:tom_hide_sample} along with the ground truth mask in the 2nd panel. 

\subsection{MeerKAT Array simulations}

The second dataset simulates the MeerKAT telescope array that  consists of 64 antennas, each  13.5m in diameter. These 64 antennas lead to 2016 independent baselines. The array has a dense core with 48 antennas located within a diameter of 1 km, with the remaining 16 spread out, giving a maximum baseline length of 8 km. 

Due to data size and computation limitations we chose 15 baselines to give the optimal trade-offs between time, frequency and baseline coverage in our data. The antennas were chosen so the baseline lengths covered scales from 100m to 8 km  and sampled all directions roughly equally. We used a MeerKAT bandwidth of 856 MHz centered at 1284 MHz and channelized at 208.984 kHz into 4096 channels. The MeerKAT feed is linearly polarized and hence measures the horizontal (H) and vertical component (V) of the electromagnetic radiation.

The MeerKAT simulations include a simplified (Airy disk beam for the HH and VV components) full-polarization primary beam model that was fitted to a Zernike polynomial representation from \cite{asad2019primary}. Neither pointing errors nor gain variations of the primary beam over time were included. Such gain variations could be considered as being factored into the bandpass time variation. A full-polarization bandpass was generated from measurements of a calibrator source with the real MeerKAT array. Time dependence of the bandpass as a whole was introduced through randomly sampling Fourier modes with periods in the range of 10 minutes to several hours and designed to mimic real data. The amplitude of the bandpass was kept within 10\% of the original amplitude. 

The astronomical source model (positions, fluxes, and source shapes) used for all MeerKAT RFI simulations comes from a combination of the SUMSS and NVSS catalogues. Spectral indices are not available from these catalogues and were therefore randomly sampled from $\mathcal{N}(-0.9, 0.3^2)$. 

The sources of RFI present in the simulations are satellites present in the L-band and the nearest 5 towns to the telescope site. The RFI originating from the 5 towns is simulated as RFI sources sitting on the horizon. RFI entering the beam from the horizon was treated like any other source where its beam attenuation is determined by its angle from the pointing centre. Multi-path effects of RFI were not taken into consideration. Each RFI source was randomly assigned a frequency range in which it was emitting. The distribution of these frequency ranges was taken such as to mimic the average RFI distribution of historical MeerKAT data, Each RFI source was then assigned a random spectral profile consisting of between 2 and 5 Sersic profiled peaks. The RFI considered in these simulations originate from man-made sources and therefore tend to be polarized. Each RFI source was randomly assigned a specific polarization (Q, U, or V) and was treated as being fully polarized.

For a radio interferometer like MeerKAT (with respect to its observing frequency and spatial distribution of antennas) many sources of RFI lie in the near-field. In these simulations near-field effects have not been taken into consideration. The open source framework (Montblanc) used to perform the RIME calculations only allows one the addition of far-field sources. 

The time dependence of each RFI source was again introduced through randomly sampling Fourier modes except with periods in the range of seconds. To obtain realistic satellite paths the Two-Line Element sets (TLEs), for each satellite were used to predict their positions for a given time and date, \citep{vallado2012two}. The output of the simulations is the complex-valued visibilities for the astronomical component, $\mathcal{A}$,  and the RFI component, $\mathcal{R}$, separately. This was done so that contaminated visibilities could be generated through a linear combination of these components and additive noise after the fact to allow for augmenting the training data, as described below.

We simulated 100 samples which each contain 800 seconds of observations for all 4 vertical (V) and horizontal (H) cross-polarizations (HH, VV, HV, and VH). We then averaged the data along the time axis in 8-second bins. The resulting data files have visibilities as a function of the following variables  (Time, Baseline, Frequency, Polarization) and of shape $(100, 15, 4096, 4)$. Each sample is 15 GB so the whole dataset is 1.5 TB in size. We split the 100 files randomly into training, validation and test set in the ratio: 40:20:40.

Despite the relatively large size of our simulated data (1.5TB)\footnote{Made up of the 15 baselines, 8s time sampling,  4 polarizations and 4096 frequency channels.} we would like more time duration in the data. To address this problem we simulated the astronomical, RFI and noise signals separately. This allows us to efficiently augment our data by randomly combining astronomical, RFI and noise data as follows:

\begin{equation}
\mathcal{D} = \mathcal{A} + \alpha \mathcal{R} + \mathcal{N}
\end{equation}
Here $\mathcal{A}$ is data for the astronomical source information,  $\mathcal{R}$ is a template RFI contaminated data, $\mathcal{N}$ is noise generated from a random complex Gaussian with standard deviation randomly chosen from a uniform distribution in the range $0.168$Jy to $0.336$Jy. $0.168$Jy is the theoretical noise for MeerKAT in a single baseline and time-frequency bin calculated from the measured SEFD of the instrument. This range therefore gives a realistic noise range for MeerKAT.  The parameter $\alpha$ controls the amplitude of the RFI contribution and was randomly chosen in the range from $2^{-10}$ to $1$ in logarithmic
bins. The resulting cumulative distribution of RFI amplitudes is shown in Fig.~\ref{fig:rfi_pdf}.

\begin{figure}
  \includegraphics[width=0.5\textwidth]{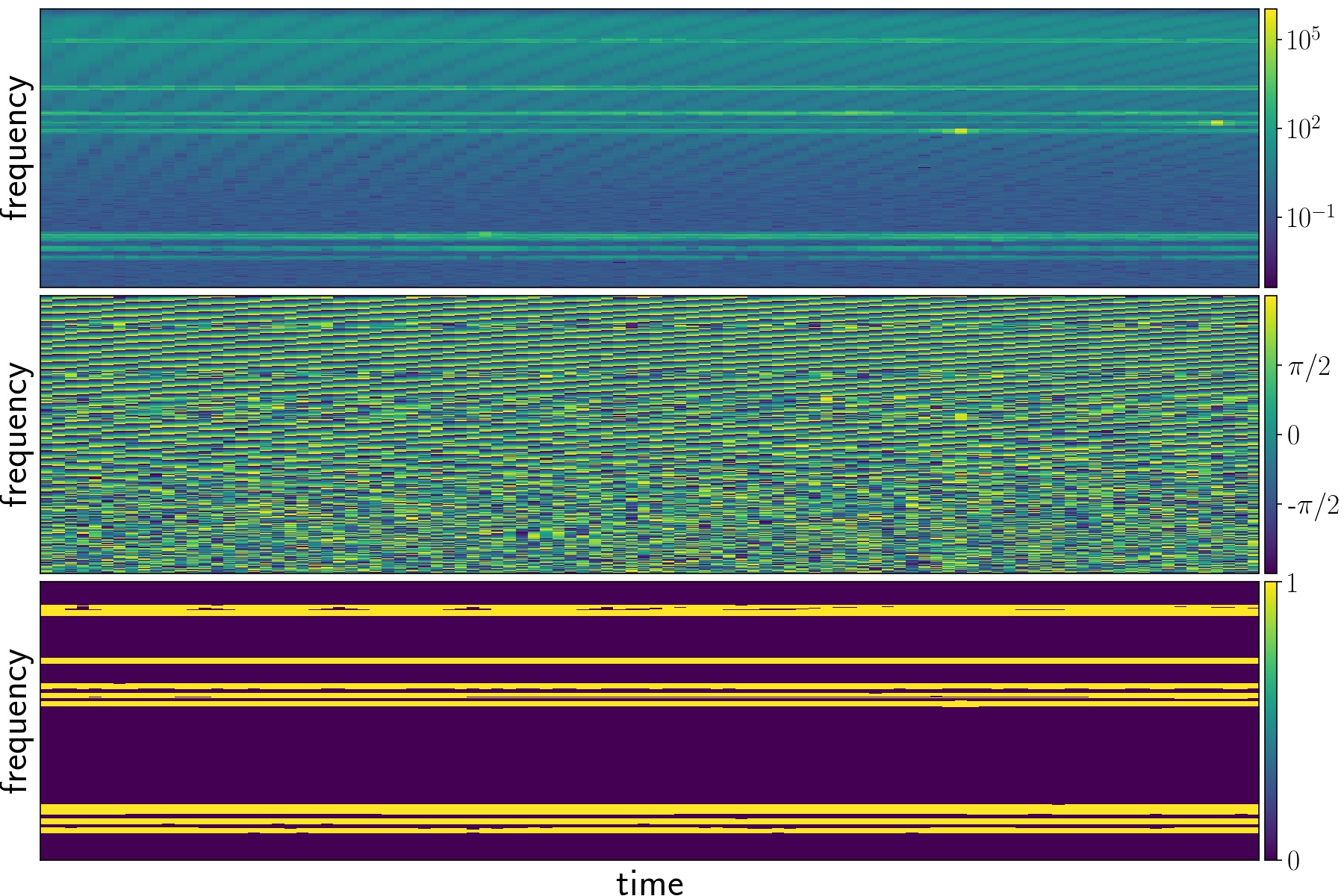}
  \caption{{\bf Example of an augmented MeerKAT simulation} for a single baseline with  $(\alpha,\sigma)=((0.1, 0.3)$. The first row is a map of absolute values of the visibility data for time vs frequency. The second row is the phase vs time and frequency (wrapped between $=\pi$ and $-\pi$) whilst the lowest panel is the ground truth mask for  $\tau_{5}= 5 \times 0.168$Jy. 
}
  \label{fig:mk_sample}
\end{figure}

Due to disk space limitations augmentation is done on the fly in memory during training and validation. The test set is augmented twice (leading to 80 $\times$ 800 s samples) and saved to disk. Because the test set is prepared only once and saved, we use the same test data for all algorithms ensuring that the details of the augmentation do not affect the comparisons. 
An example of an augmented MeerKAT simulation sample (800 seconds, 4096 frequencies) is shown in Fig. \ref{fig:mk_sample}.

\subsection{Real \kat\ observations}

Our third dataset comes from 12 hours of actual observations from the Karoo Array Telescope (KAT-7).
\kat\ was an engineering prototype for the MeerKAT telescope array, but lead to its own scientific contributions ~\cite{foley2016engineering}.

The data includes all 7 antennas that lead to 42 baselines observed in the summer of 2016. More detailed information about the dataset is available in \cite{heald2016neutral}. The  RFI in the \kat\ data are
manually flagged by experts and we used these manually obtained RFI 
masks as ground truth. We note that the hand flagged mask will of course not be perfect but is the best approximation to the truth that we have for real data.

The purpose of using the \kat\ data is to show that, given a relatively small amount of human-flagged data, our algorithm is able to transfer from simulations to real data with high precision, delivering results at a speed comparable to or better than AOFlagger and similar non-deep learning algorithms. This is an important step since otherwise one would have to rely on simulated RFI data being highly realistic. Even in this case there would be a nagging concern that elements of the full telescope not captured in the simulations might lead to significant degradation of the performance of the RFI flagger when applied to real data.

\subsection{Ground truth maps}

The goal of RFI flagging and excision is to remove all pixels that are contaminated by RFI beyond a threshold determined by the desired image quality and goals of the science: hence we address this as a classification problem and output a binary mask~\footnote{Finding the optimal threshold is specific to the science under consideration. One type of science may require high purity, another may prefer to have more data but with higher levels of RFI contamination. Determining this would require a full optimisation based on the underlying science; see e.g. \cite{eyeswideopen,eyesII}.}.

\begin{figure*}
  \centering
  \includegraphics[width=0.45\textwidth]{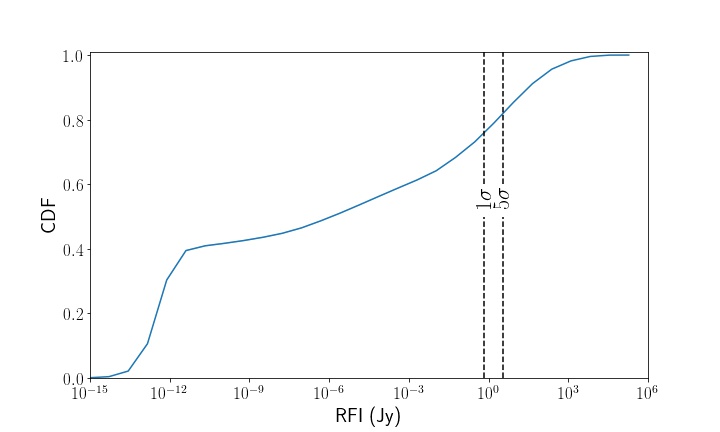}
  \includegraphics[width=0.45\textwidth]{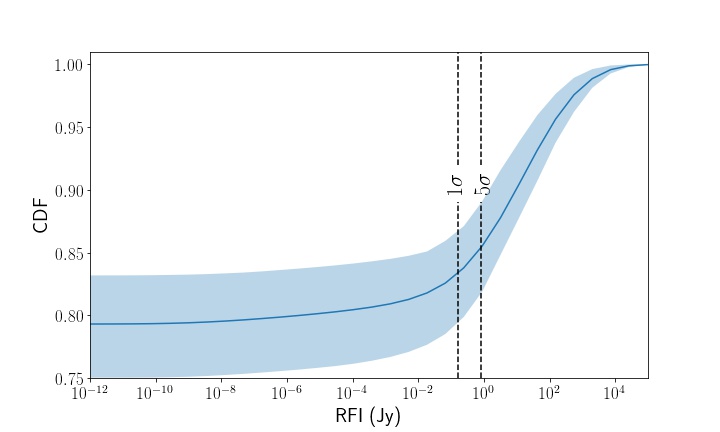}
  \caption{{\bf Average Cumulative Distribution Functions (CDF) of the daily averaged RFI amplitude for the HIDE (left) and MeerKAT (right) simulations}. The $1\sigma$ and $5\sigma$ thresholds are shown. We choose $5\sigma$ as our primary threshold, leading to $41\%$ and $16\%$ of pixels being masked as RFI pixels in the HIDE and MeerKAT simulations respectively.  The shaded areas show 95\% confidence levels for both plots. The HIDE simulations show very small variation in the CDF resulting in very narrow error bars.
  }
  \label{fig:rfi_pdf}
\end{figure*}

As we use supervised learning algorithms for pixel classification we need ground truth binary maps, $\mathcal{M}$, to learn from. In contrast, the SDP Online Flagger does not need ground truth data to run out of the box, although we do use the ground truth data to optimise the \aoflagger.   

Since the RFI amplitude varies continuously over the pixels we create binary ground truth maps by choosing an RFI threshold $\tau_{n}$. For this purpose we used  $\tau_{n}=n\sigma$ where $\sigma$ is the signal to noise ratio and $n = 1$ or 5, is an integer which controls the RFI amplitude, $\mathcal{R}$, needed to classify a pixel $ij$ as either clean or contaminated:

\begin{equation}\label{eq:mask}
    \mathcal{M}_{ij} = 
     \begin{cases}
       1   & \mathcal{R}_{ij}   >  \tau_{n}\\
       0   & \mathcal{R}_{ij} \leq \tau_{n}\\
     \end{cases}
\end{equation}

Making  $\tau_{n}$ too large results in too few masked pixels which are very easy to find. On the other hand, for very small $\tau_{n}$, most pixels are masked and again we end up with a highly unbalanced dataset. Further, because of the large noise level, there is little signal to learn from. 

We consider two values: $\tau_{1}=1\sigma$ and $\tau_{5}=5\sigma$. Since this value changes for each dataset, the thresholds $\tau_n$ will also vary from dataset to dataset even for fixed $n$. We choose $5\sigma$ for our main results. Results for $1\sigma$ case are shown in the appendices but are qualitatively similar. 

The RFI distribution for both HIDE and MeerKAT simulations are shown in Fig. \ref{fig:rfi_pdf}. For the HIDE simulations $\sigma=0.670$Jy  while for MeerKAT simulations $\sigma=0.169$ Jy. The $\tau_{5}=5\sigma$ threshold on average masks $41\%$ of pixels in the HIDE simulations and $16\%$ of the pixels in the MeerKAT simulations. Choosing $\tau_{1}=1\sigma$ in contrast, leads to $46\%$ and $17\%$ of pixels being  masked in the HIDE and MeerKAT simulations on average.


\section{New Deep RFI Algorithm: \rnet}\label{model}

Almost all supervised learning problems are optimizations of some very flexible and non-linear model to return the desired output. The optimization objective is normally a loss function which captures how close the model is to the desired output. 
Neural networks, deep learning and specifically Convolutional Neural Networks (CNN) show very promising results in different problems among other machine learning algorithms.

\begin{figure*}
	\centering
	\includegraphics[scale=0.45]{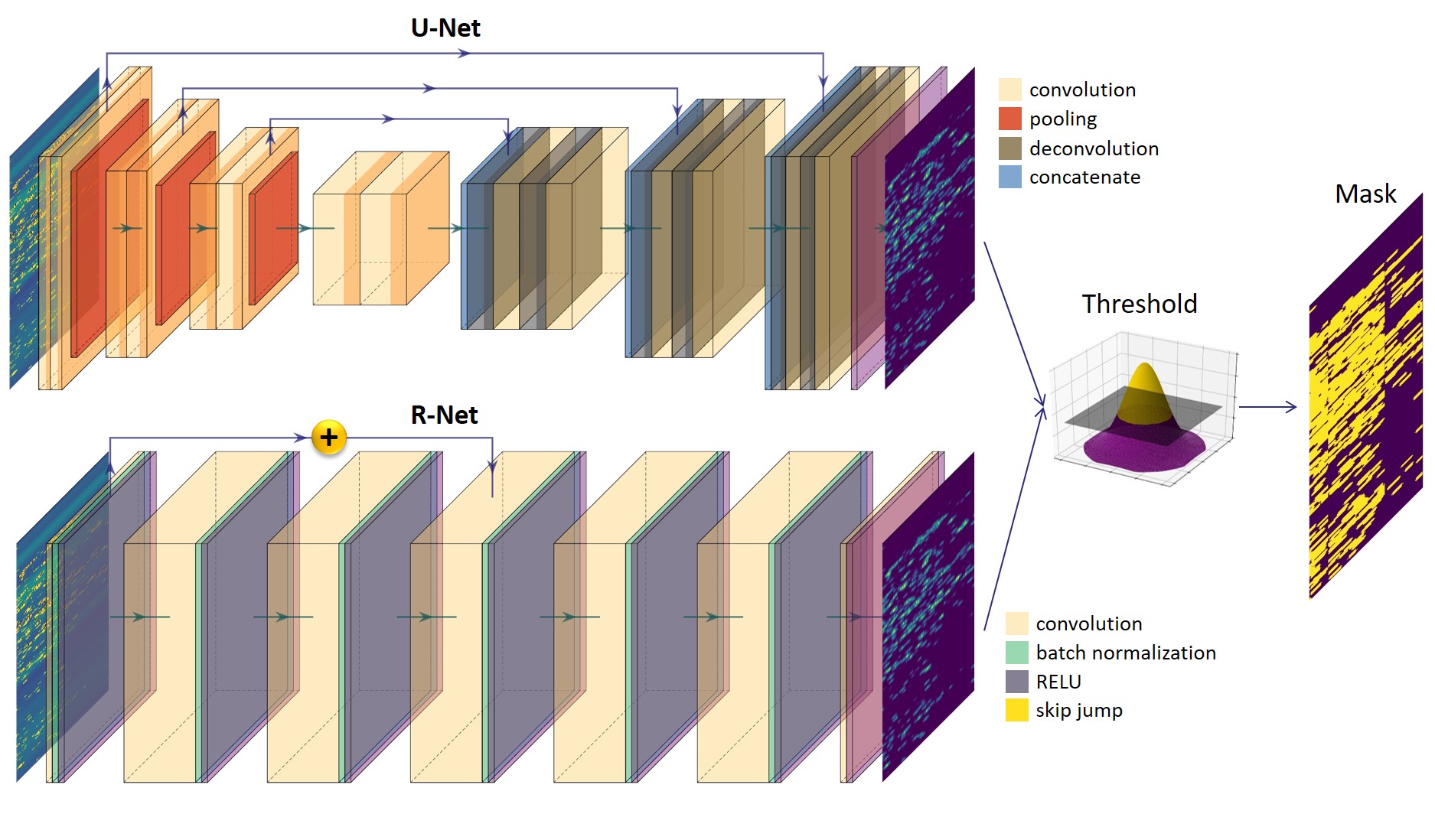}
	\caption{{\bf Schematic views of the \unet\ architecture (top) and our new \rnet\ algorithm (bottom)}. Both algorithms predict RFI probabilities for each pixel in the map. The final binary mask is produced by passing the output maps through a thresholding process which is one of the hyperparameters for the algorithms. }
	\label{fig:architecture}
\end{figure*}

In this study, we use a CNN based model to do a binary classification of every pixel of the time-frequency images given the pixel and those around it. The output is a binary mask image of the same size. This is similar to segmentation problem in computer science \citep{badrinarayanan2017segnet,ronneberger2015u,badrinarayanan2017segnet,zhang2017image,alom2018recurrent}.

Our proposed network, which we call \rnet,  is inspired by the residual network architecture (ResNet) (see \cite{he2016deep,szegedy2017inception}). Deeper neural networks can be significantly more difficult to train (e.g. due to the vanishing gradient problem, see \cite{hochreiter2001gradient}). 
The ResNet was introduced to ease the training process of the deeper layers by adding shortcuts between some layers (e.g. up to 152 layers). 
We use one such shortcut to build our architecture, constructed using sequences of 2D convolutional layers without any reduction in the size of the input. For computational reasons we did not consider more than six layers and hence did not need more than one skip/shortcut.  

\rnet\ is very simple in terms of its architecture since it consists only in convolutional layers with zero paddings to conserve the size of the image as it moves through the layers. After each convolutional layer (except the last one) we find that inserting batch normalization layer and using RELU activation functions to be optimal. The other hyperparameters of such an architecture are the number of layers, kernel size, number of filters, location of shortcut(s) and activation function.  Searching the hyperparameter space is very time consuming and limited by available computing resources: our exploration of the potential hyperparameters was guided by intuition and trial and error on the validation set. 

All of the hidden layers contain $12$ filters and the kernel size is always $5\times 5$. \rnet\ does not change the input size through layers so any number of hidden layers works;  we try 3, 5, 6 and 7 number of layers.
A shortcut (residual network) connection directly transfers information from earlier layers to the deeper layers to avoid the degradation problem \citep{he2016deep}. We used no shortcut connections for the 3 layer architecture, one for the 5 and 6 layers architectures and 2 shortcut connections for the 7 layer architecture.  Our initial results show the performance of the 5 and 6 layers architectures (hereafter \rnet5 and \rnet6) provide the best performance  better in term of AUC.  

The output layer is always a $1\times 1$ convolution layer which aggregates all $12$ filters into a one channel image. The output images are normalised  between 0 and 1 and, roughly speaking, each pixel value shows the probability of being RFI-contaminated. 
It is possible to turn the output into a binary mask by applying a threshold and compared to the ground truth whose pixels are either 0 or 1 representing clean and RFI-contaminated pixels respectively. We try mean square error (MSE) and binary cross entropy as loss functions and find that MSE performs better. This suggests reconsidering RFI problem as a regression problem: an idea we present in a future work.

\section{Comparison of Algorithms}

To fully evaluate the performance of \rnet\ we compare it against two existing RFI algorithms: the deep \unet\ algorithm, and the standard \aoflagger.  

\subsection{\unet}

\unet\ is a CNN-based approach (see \cite{ronneberger2015u}) first used for RFI flagging  by \cite{akeret2017radio} in the context of single dish experiments.
It includes convolutional layers that are usually used in image classification
problems since they build a conceptual hierarchy that can extract different kinds of features through down sampling. 
The \unet\ architecture first contracts and then expands the input images back to the original size using deconvolutional
(see \cite{zeiler2010deconvolutional}) and upsampling layers. Roughly speaking, the bottleneck shape, like autoencoders, are supposed to force the CNN to learn the most important features and avoid overfitting.

The \unet\ architecture has been used for RFI detection in single dish simulations  \citep{akeret2017radio}, where it showed improved performance over a standard non-machine learning algorithm. You can see a schematic illustration of both the \unet\ and \rnet\
architectures in Fig.~\ref{fig:architecture}.
We use both \unet\ and \aoflagger\ to compare with \rnet.

\subsection{The MeerKAT SDP Online Flagger}
The SDP Online Flagger\footnote{https://github.com/ska-sa/katsdpsigproc} (hereafter called the \sdp)
 is the default RFI flagger used by the MeerKAT Science Data Processing (SDP) pipeline. It is based on the classic AOFlagger\footnote{https://sourceforge.net/p/aoflagger/wiki/Home/}, a C++ RFI flagging algorithm that uses  morphological features \citep{offringa2010post}. AOFlagger runs on a two-dimensional data array of time and frequency visibilities to flag and mask RFI signals and has been used in a number of interferometric telescope arrays, including LOFAR, WSRT, VLA, GMRT, ATCA and MWA, and single-dish telescopes such as the Parkes and the Arecibo telescopes.

In order to make the comparison to \rnet\ as fair as possible, we optimize \aoflagger\ performance by varying its various hyperparameters. Due to computational limits, and after experimentation, we choose to vary \textit{outlier-n-sigma},  $\sigma_O$, and \textit{background reject}, $R_b$, as the most significant hyperparameters.  In appendix \ref{app:aoflagger} we show in Fig (\ref{fig:sdp_compare})
the performance of \aoflagger\ when $\tau_{n}$, $\sigma_O$ and $R_b$ vary. 


\begin{table*}
\begin{center}
\begin{tabular}{lllllll}
\toprule
Algorithm & HIDE-$5\sigma$ & HIDE-$1\sigma$ & MeerKAT-$5\sigma$ & MeerKAT-$1\sigma$ &  KAT7 \\
\midrule
R-net5    &  0.96 &  {\bf 0.96} &  0.77 &  {\bf 0.76} &  0.79 \\
R-net6    &  {\bf 0.97} &  {\bf 0.96} &  {\bf 0.78} &  0.68 &  {\bf  0.85} \\
U-net16   &   0.78 &  0.80 &  0.37 &  0.35 &    -  \\
U-net32   &  0.80 &  0.81 &  0.39 &  0.38 &    -  \\
SDP Flagger &  0.69 &  0.71 &  0.59 &  0.62 &  0.63 \\
\bottomrule
\end{tabular}
\caption{{\bf F1-scores for the various different algorithms and datasets} considered in this paper. The best results for each dataset are shown in bold. The \rnet\ models outperform all the \unet\ and  \aoflagger\ models by a considerable margin.
 \label{table:f1score}}
\end{center}
\end{table*}

\section{Training}

For all simulations, the datasets were split into training, validation and test sets. We use the validation set for hyperparameter optimization and the results are reported on the test set.  

Since \rnet\ and \unet\ architectures are fully convolutional, it is possible to train the models on smaller  frequency-time images and use the final model on larger datasets. In fact, this is one advantage of using neural networks: they are expandable to run on GPU. For training window size we used all frequencies and a limited the size of time steps to $W_t$, which is treated as a hyperparameter: large $W_t$ leads to memory overload while small $W_t$ leads to poor performance. Therefore we maximise $W_t$ given memory constraints.  

We use the cross-entropy loss function for the \unet\ architecture and MSE for \rnet. We used the RMSPropOptimizer optimizer with the learning rate initiated to 0.2 and decreasing exponentially/linearly
for \unet/\rnet\ and the weights are initiated using the truncated normal distribution. 

One notable point on training  \rnet\ is that for the MSE loss function models can easily be trapped in a local minimum and predict a constant value for all pixels. This problem is more dramatic when the data labels are unbalanced, as happens in the MeerKAT simulations. We call this the ``dead model'' problem.
To address this issue, we use two recommended techniques in the \href{Ngene}{\footnote{https://github.com/vafaei-ar/Ngene}} package
The first is model resuscitation where the model gets checked frequently during the training process.  The constant value returned in dead models, is usually close to the average of the labels. To check this condition, one can use the output's standard deviation at a certain frequency while the model is training. Model resuscitation can be simply done by re-initiation of the weights, which suffices for our purposes.

The second issue is overfitting, where a model easily learns noise and focuses on irrelevant patterns as information. The problem gets worse when the number of trainable parameters is large. There are plenty of different proposals like data augmentation, regularization, dropouts and etc. to address this problem. 

The core ML models are flexible and contain trainable parameters, increasing the  chance that the model overfits. We exploit this tendency to overfit to initialize the model and train the neural networks given only a few data points and allow the CNN model overfit. This usually lets the neural network use the overfitted weights as good initial values for the whole training set. 

\rnet\ was developed using {\em Ngene} \footnote{https://github.com/vafaei-ar/RFI-mitigation}. Both \unet\ and \rnet\ are trained on NVIDIA Tesla P100 12GB GPUs until the average loss does not improve for 50 epochs.  


\section{Results}
\label{results}

To compare \rnet\ against the other two reference algorithms (\unet\ and \aoflagger) we consider several metrics.
The first is the AUC, corresponding to the Area Under the True Positive Rate (TPR)-False Positive Rate (FPR) curve,~\citep{bradley1997use}, which is usually a good metric for binary classification problems.

However, since the datasets we consider can be significantly unbalanced (with few RFI-dominated pixels) the usual classification metrics may give misleading results. We therefore also compute the Precision-Recall curves, F1-score and Matthews Correlation Coefficient (MCC) which are better metrics for unbalanced data; see e.g.  \cite{davis2006relationship}. For more details see appendix \ref{app:metrics} or \cite{chicco2020advantages}. F1-scores are shown in table \ref{table:f1score}. The 5 and 6-layer versions of \rnet\ outperform all other variants of \unet\ and \aoflagger\ significantly in all metrics. Note that it is often in the precision-recall plots that \rnet\ significantly outperforms the other algorithms.   

For all datasets, we varied the \aoflagger\ hyperparameters $\sigma_O$ and $R_b$. The results are shown as the point cloud in all figures.  The effect of $\sigma_O$ and $R_b$ on fall-out, recall, precision are shown in figure \ref{fig:sdp_compare}.

\begin{figure*}
	\centering
	\includegraphics[scale=0.4]{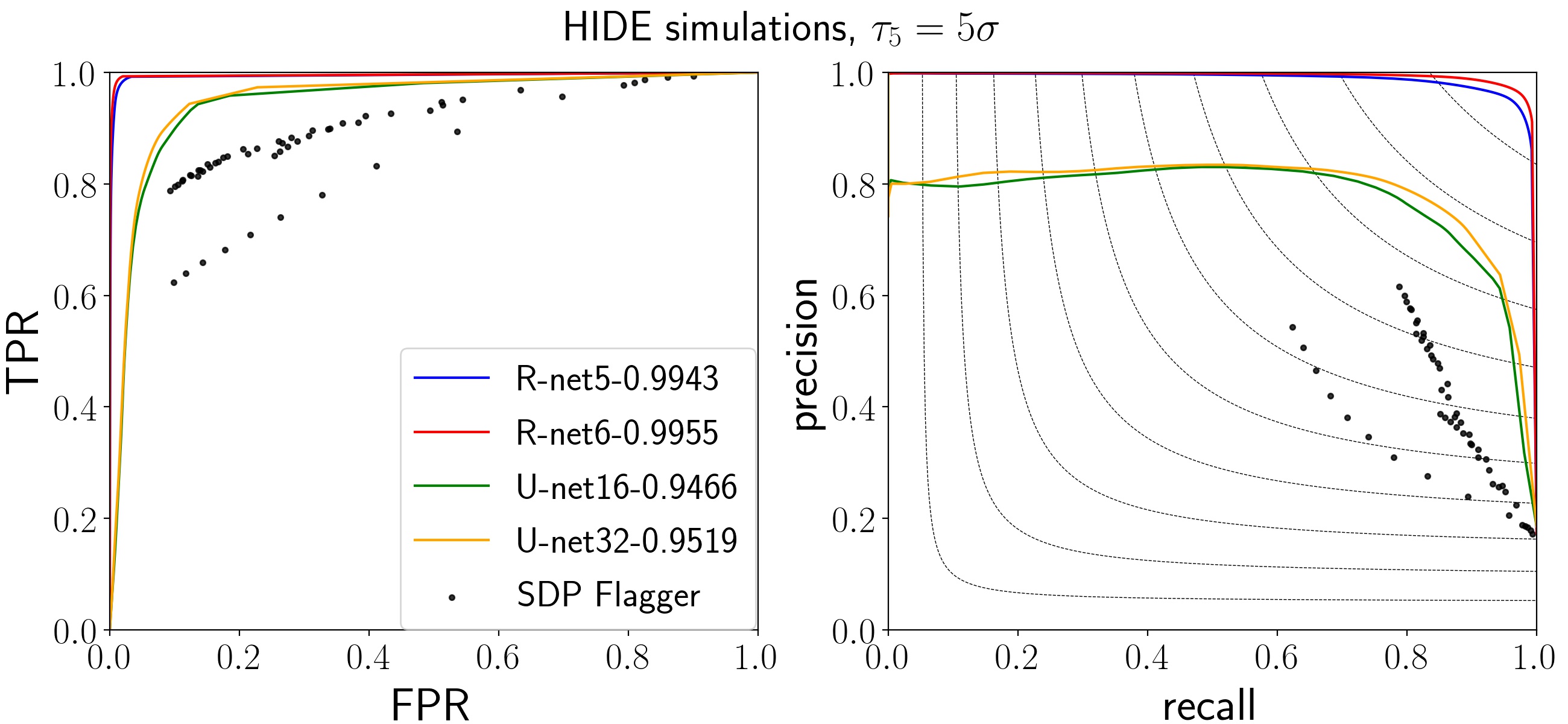}
	\caption{{\bf Performance on single-dish simulations - } \rnet\ is almost perfect and significantly outperforms both \unet\ and the \aoflagger\ when trained on single-dish HIDE simulations when predicting the RFI mask at an RFI threshold of $5\sigma$.  The legend shows corresponding AUC values for \rnet\ and \unet. Dots are the \aoflagger\ results using different hyper parameters. The \aoflagger\ performs poorly for all values on the precision-recall plot. Contours denote iso-F1 scores. The plots for $1\sigma$ RFI threshold are shown in the appendix, and are qualitatively similar.}
	\label{fig:hide-cnn1}
\end{figure*}

\begin{figure}
	\centering
	\includegraphics[scale=0.38]{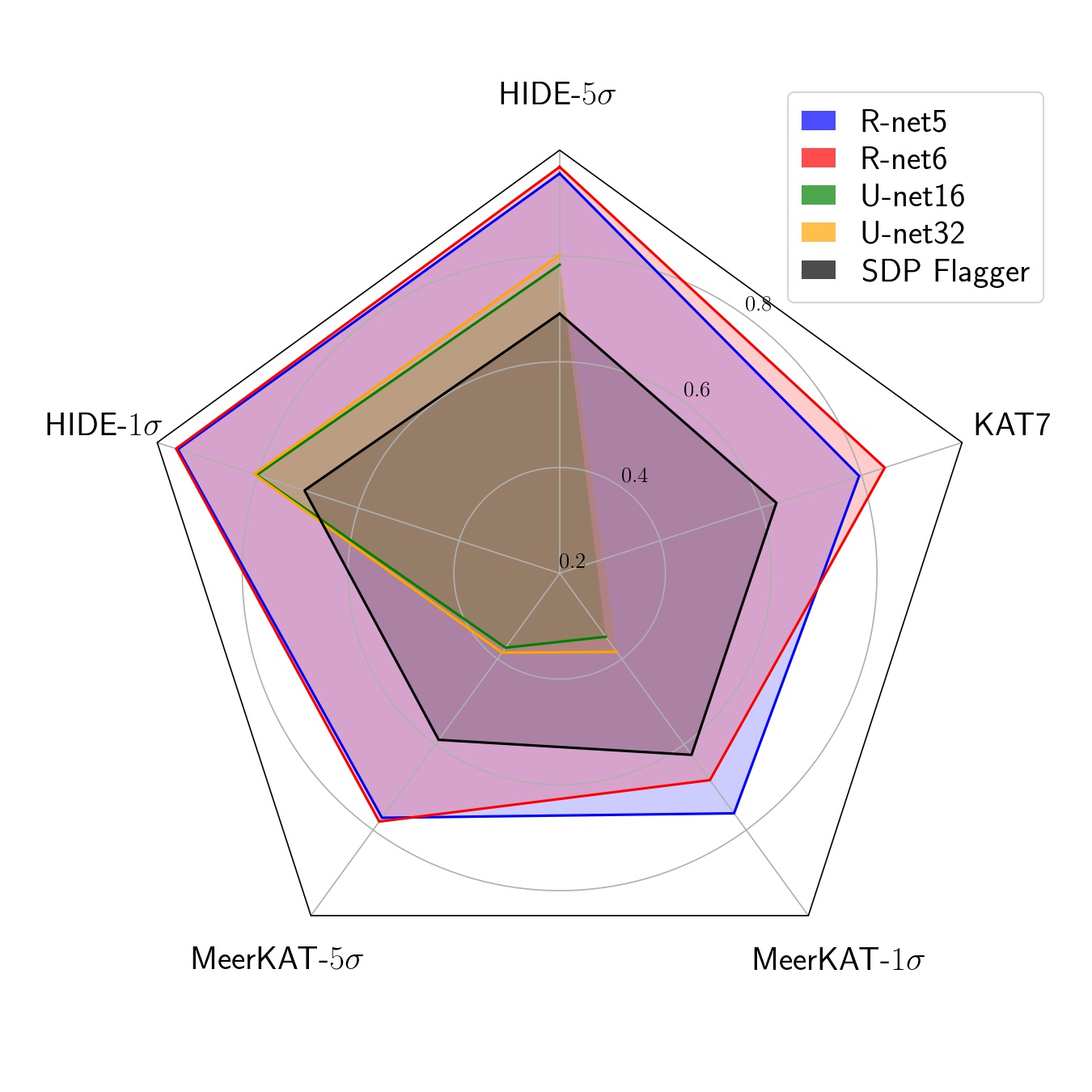}
	\caption{Starplot comparison between F1 scores of all the algorithms for all the datasets and different thresholds (e.g. HIDE-$5\sigma$ corresponds to the results where the threshold is $\tau_5$.) The center corresponds to a score of zero and the outer contour to a perfect score of 1. \rnet\ outperforms all other algorithm variants as it encloses all other algorithms on all datasets and threshold. Note that \unet\ was not included for the transfer learning task on the \kat\ dataset.}
	\label{fig:starplot_f1}
\end{figure}

A summary of the performance of all the algorithms on all the datasets and all thresholds is shown in Fig. \ref{fig:starplot_f1} in terms of the F1-score. While this is only one metric it does illustrate the power of \rnet. We now discuss performance of the algorithms on each dataset separately.

\subsection{HIDE simulations}

We split the 13 months of data into 11 months for training, 1 month for validation and 1 month for the test set. We chose $W_t=400$ (the width of the time window for training). 

The best \aoflagger\ F1-score results for $5\sigma$ and $1\sigma$ thresholds were $0.69$ and $0.71$ .
The best result for the \unet\ architecture was achieved by using 32 features in the latent layer and achieved (AUC,F1-score) = (0.95,0.80)  for $5\sigma$ and (AUC,F1-score)=(0.94,0.81) for the $1\sigma$ threshold.

\begin{figure*}
	\centering
	\includegraphics[scale=0.35]{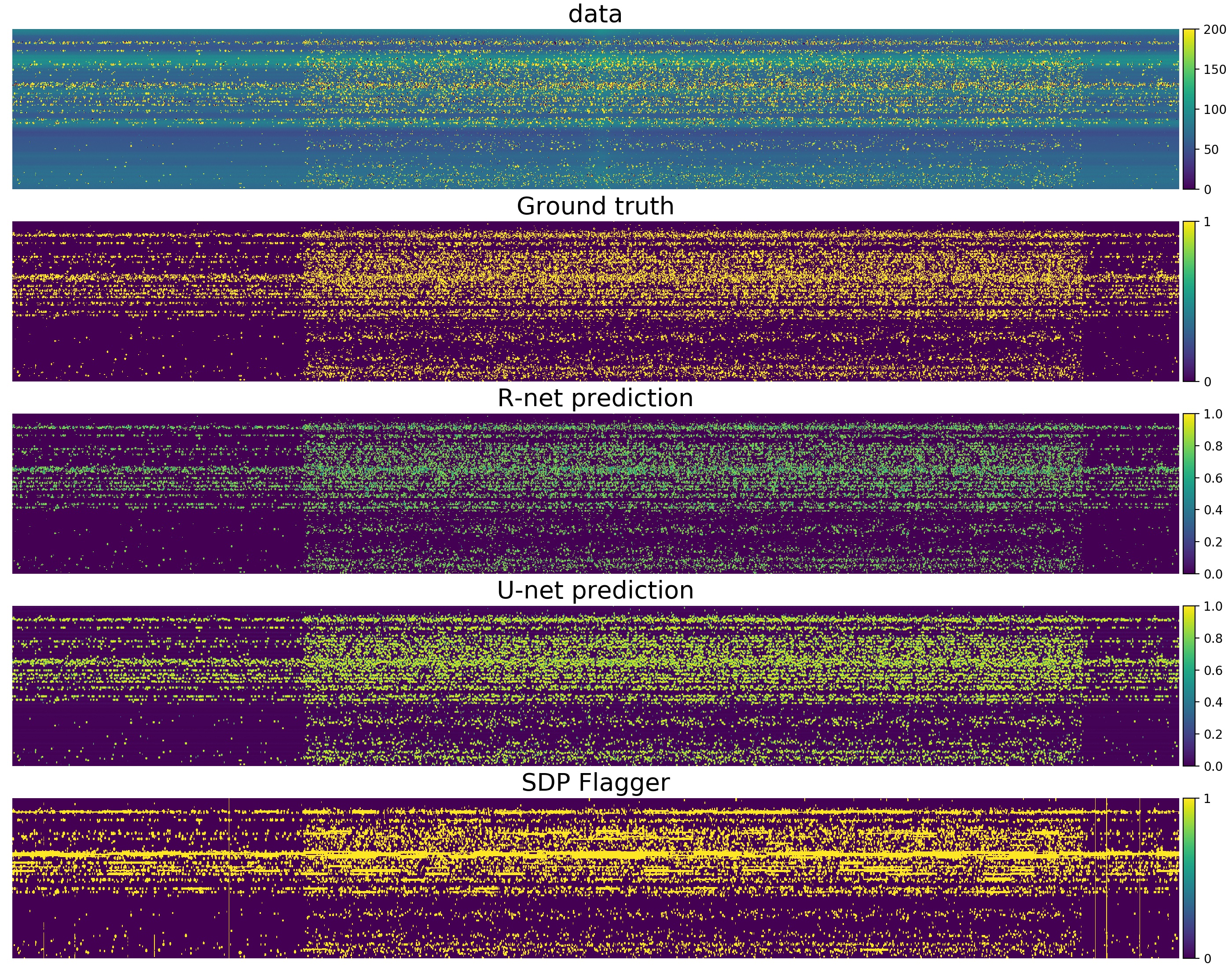}
	\caption{{\bf One day of HIDE simulations along with the results of the various flaggers.} The first row shows the observed data, clipped between $0$ and $200$. The second, third and forth rows show ground truth, \rnet, and \unet\ probability outputs. The final row is the output of \aoflagger. The axes are time and frequency similar to Fig. \ref{fig:mk_sample}}
	\label{fig:tom_hide_sample}
\end{figure*}

For the \rnet\ architecture the best results for $\tau_{5} \equiv 5\sigma$ threshold is (AUC,F1-score)=(0.99,0.97) and for $\tau_{1}=1\sigma$ is
(AUC,F1-score)=(0.99,0.96). Using 6 layers results in slightly better results over 5 layers. \rnet\ significantly outperforms both \aoflagger\ and \unet\ and returns almost perfect results. A sample of one day simulated data together with results from the various algorithms is shown  in Fig. \ref{fig:tom_hide_sample}. The ROC and precision-recall curves for the HIDE simulations are shown in Fig. \ref{fig:hide-cnn1} and \ref{fig:hide-cnn2}.

\subsection{MeerKAT simulations}

The MeerKAT dataset consists of 100 files with 800 seconds of simulated data. 40 files are augmented in memory and used for training, while 20(40) files are augmented 3 times for validation (test) sets. We averaged the time axis every 8 seconds and chose $W_t=50$ for the width of the time window.

The best \aoflagger\ F1-scores were $0.69$ and $0.71$ for the $5\sigma$ and $1\sigma$ thresholds. The best result of the \unet\ architectures was achieved for 32 features in the latent layer achieving (AUC,F1-score)=(0.62,0.39)  for $5\sigma$ and  (AUC,F1-score)=(0.59,0.38) for a $1\sigma$ threshold.

\begin{figure*}
	\centering
	\includegraphics[scale=0.4]{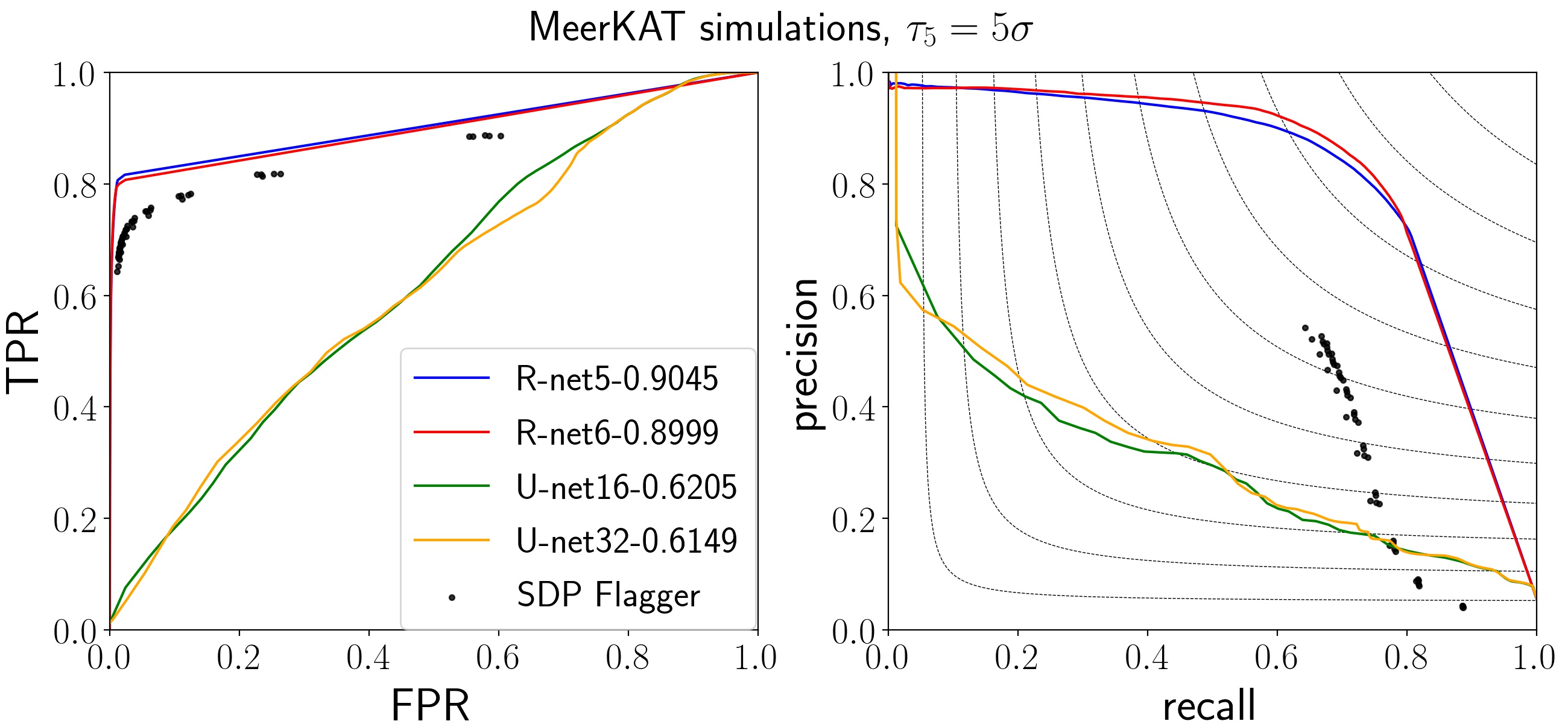}
	\caption{{\bf Performance on MeerKAT simulations -} \rnet\ significantly outperforms both \unet\ and the \aoflagger\ trained on MeerKAT simulations using only the absolute value of the visibility data (RFI threshold of $5\sigma$). The legend shows corresponding AUC values for \rnet\ and \unet. Note the significant drop in performance of all algorithms relative to their corresponding results on the much simpler HIDE data. Dots show the \aoflagger\ results using different hyper parameters. Although the \aoflagger\ does relatively well on the TPR-FPR curve it performs poorly for all values on the precision-recall plot. Contours denote iso-F1 scores. The equivalent plots for $1\sigma$ RFI threshold are shown in the appendix, and are qualitatively similar. 
	In Fig.~\ref{fig:mkph-cnn1} we show the equivalent plot with phase information included, demonstrating that uncalibrated phase information adds no additional significant value to \rnet.
	}
	\label{fig:mk-cnn1}
\end{figure*}

For the \rnet\ architecture using only the amplitude of the visibilities,  the best result was (AUC,F1-score)=(0.90,0.77) and  (AUC,F1-score)=(0.87,0.76) for the $5\sigma$ and $1\sigma$ thresholds respectively, with slightly better results from the 5-layer architecture. \rnet\ significantly outperforms  all \aoflagger\ and \unet\ configurations with respect to the investigated metrics. TPR-FPR and precision-recall curves for the MeerKAT simulations are shown in Fig. \ref{fig:mk-cnn1} and \ref{fig:mk-cnn2} for the two different thresholds.

We also investigated adding phase information in addition to the amplitude of the visibilities as one of the channels of the input layer of the similar architecture and the results do not significantly improve the performance of the algorithm, as shown in Fig.~\ref{fig:mkph-cnn1}. This is likely due to the fact that the phase is not calibrated.

\subsection{Transfer Learning for \kat}

We have so far only considered the performance of the algorithms on simulated data. The concern with this is that the resulting ``best" algorithm might actually perform badly on real telescope data due to the deficiencies in the simulated data. 

This could be circumvented if one had very large amounts of accurately labeled real data by simple training an \rnet\ model from scratch. However, large amounts of human-labeled, real data are not available (especially for large arrays with many baselines). Further, human labels are imperfect, especially in regions where the RFI contamination is fairly week, and therefore will anyway contain errors.  

To deal with this catch-22 problem we investigate the use of transfer learning applied to the \rnet\ model trained on the MeerKAT data. {\em Transfer learning} \citep{tan2018survey} in this context means we freeze the weights in all of the layers before the shortcut leaving only the weights in the last two layers to be trained. Freezing the lower layers means we retain the key ability of the network to recognise relevant RFI features while leaving only relatively few parameters that need to be learned to adapt to the real data. As a result we only need a fairly small amount of real data, which also bounds the error introduced by the imperfections in the human labelling.  

\begin{figure*}
	\centering
	\includegraphics[scale=0.4]{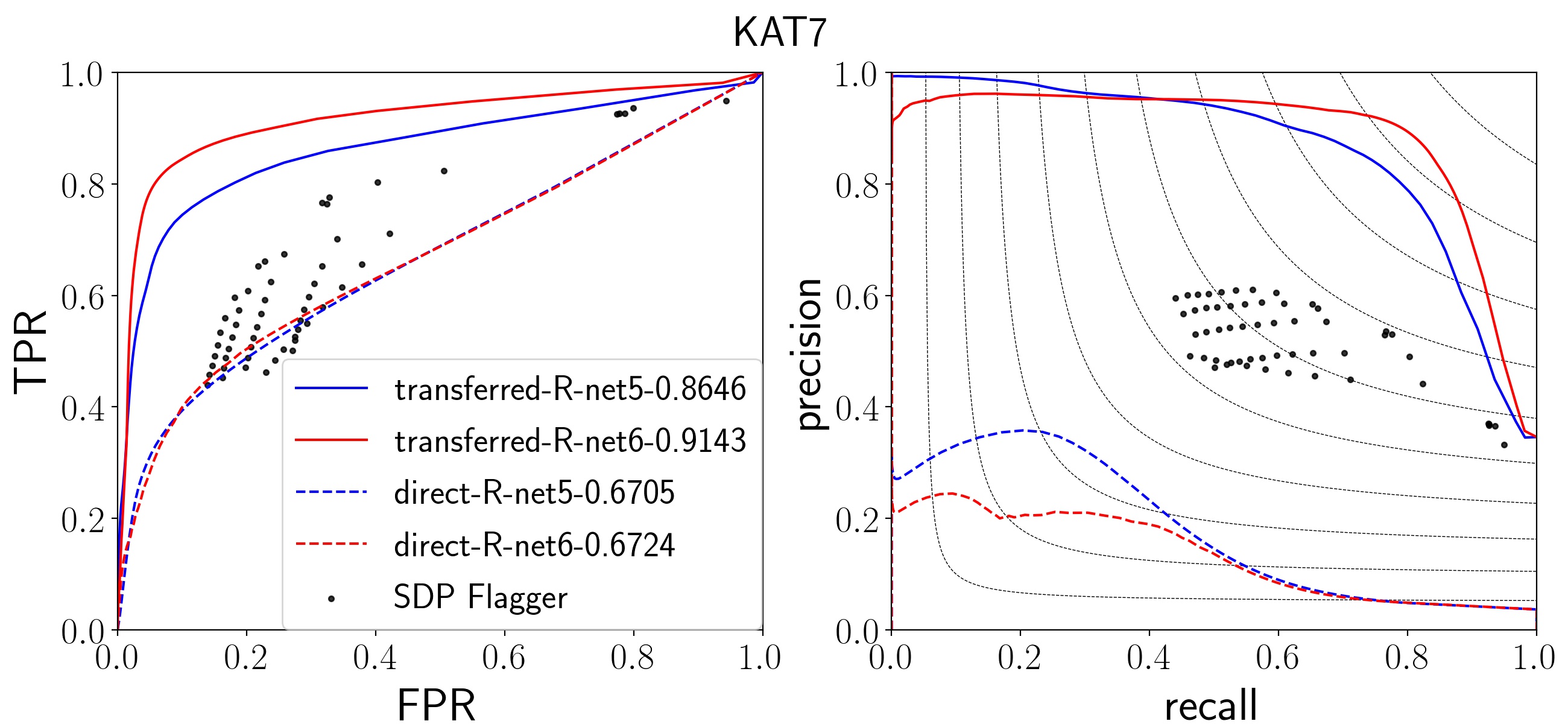}
	\caption{{\bf Transfer learning from MeerKAT simulations to real \kat\ data -} \rnet\ performance  on the \kat\ hand-flagged dataset using the best MeerKAT model (dashed lines; ''direct") and the same models after transfer learning with a small amount of real \kat\ data (solid lines; ''transferred''). Transfer learning massively improves the performance of the model. Results are averaged over 4 polarizations and are averages of 4-fold cross validation. Points indicate the various \aoflagger\ results for different hyperparameter combinations of $(\sigma_O$,$R_b)$ values. Note that at a recall of about 0.8 the best transferred \rnet\ model has approximately double the precision of the best \aoflagger\ model. Note also that there were no parameter values for the \aoflagger\ that had an FPR $< 0.1$, where as this can be tuned with the deep learning models. Contours denote iso-F1 scores. }
	\label{fig:kat7}
\end{figure*}

In this case we applied transfer learning to the human-labeled \kat\ data described earlier.  For our results we use average performance on 4-fold cross-validation. A comparison of  \aoflagger\ and \rnet\ with transfer learning is shown in Fig. \ref{fig:kat7}. \rnet\ with transfer learning on the \kat\ data actually achieves a slightly higher AUC score than on the MeerKAT simulations (0.91 vs 0.90). Even more impressive, the best MeerKAT-trained model only achieved an AUC of 0.67 when applied directly on the \kat\ data (dashed lines in Fig. \ref{fig:kat7}), showing that the transfer learning had a massive impact. To illustrate how little data was used in the transfer learning, the best \rnet\ model trained from scratch on the \kat\ data alone was only able to achieve an AUC of 0.55. While performance between the 5 and 6-layer \rnet\ models was almost indistinguishable in earlier tests, the 6-layer \rnet\ model performed significantly better in the transfer learning task, perhaps showing that the extra flexibility provided by the extra layer could be exploited efficiently.  

A summary of the results in terms of F1-score and TPR-FPR curves are shown in table \ref{table:f1score} and Fig. \ref{fig:starplot_f1}. Both \rnet\ configurations do very well.

As a result we conclude that transfer learning is a very promising approach to fine-tuning algorithms trained on simulated data. Given that the \kat\ data is very different in frequency and angular resolution to MeerKAT, the success of transfer learning here also shows that it can also be applied to fine-tuning algorithms trained on data from a different telescope.

\subsection{Performance comparison}

An algorithm that is accurate but extremely slow to evaluate on new data is of limited use in future real world applications such as the SKA which will be dominated by huge incoming data rates. We therefore  also evaluated the  computational cost of running each trained algorithm on test data; with the results shown in table \ref{table:n_variables}. 
The comparison is performed between different algorithms in terms of the number of trained variables in the case of \rnet\ and \unet,
execution time and number of FLOPs measured on a 3.3GB test dataset.  
One can see the number of trainable variables for \rnet\ is less than \unet\ by a factor $\approx 10$. 
As a result there is a higher chance of overfitting in \unet\ models. 
The number of FLOPs required by  \rnet\ is larger than \aoflagger\ and \unet\ but
despite this fact, the \rnet\ execution time is actually slightly smaller than the other methods because the \rnet\ architecture contains  convolutional layers that are all the same size and hence operations are more efficiently handled in parallel.

\begin{table}
\begin{center}
\begin{tabular}{|l||c|c|c|}
\hline
Model         & variables & exe. time(s) & FLOPs \\ \hline\hline
\aoflagger    &   -       &        31    & 6.8T  \\ \hline
\rnet-5        & 11473    &        18    & 8.3T  \\ \hline
\rnet-6        & 15181    &        24    & 11.9T \\ \hline
Unet-3-16     & 116770    &        49    & 4.0G  \\ \hline
Unet-3-32     & 465986    &        53    & 4.0G  \\ \hline
\end{tabular}
\caption{Number of trainable variables, FLOPs and execution time for the different 
algorithms. The CNN-based algorithms used a Tesla P100, 12GB GPU and \aoflagger\ 
used 6 CPU cores.
\label{table:n_variables}}
\end{center}
\end{table}


\section{Conclusion}
\label{conclusion}

We describe a new ResNet-style  convolutional neural network
algorithm for Radio Frequency Interference (RFI) flagging. We test this algorithm on both single-dish and realistic
interferometric telescope array RFI simulations showing that it significantly outperforms the current state-of-the-art algorithms (both  \unet\  and the modified version of AOFlagger currently used in the MeerKAT data reduction pipeline). 

The default \rnet\ algorithm was trained on the magnitudes of the complex visibilities. We did also explore the use of the phase to aid in RFI flagging but found that uncalibrated phase information did not lead to any additional improvements over the magnitudes alone, whether they were included by splitting the visibilities into real and imaginary components or into magnitude and phase.  

Finally we show that models trained on simulated MeerKAT data can be very efficiently fine-tuned via transfer learning to provide state-of-the-art results on real data flagged by humans from a very different telescope; in this case the \kat\ array. Transfer learning involves retraining the last few layers of the existing model on the real data. This process boosted average performance from an AUC of 67\% to 91\%. This success suggests that transfer learning will be a powerful and exciting tool for RFI flagging as we move into the era of the SKA.

\section*{Acknowledgments}
 
The authors thank Michelle Lochner and Ethan Roberts for insightful discussions and comments and the anonymous reviewer for useful comments. AVS thanks Golshan Ejlali and Anke Van Dyk for useful discussions. We thank the members of the SARAO Data Science and SDP teams and the RFI Working Group for discussions. CF thanks Ben Hugo, Simon Perkins and other members of the SARAO RARG team for useful discussions around radio interferometry simulations. The numerical computations were carried out on SARAO facilities, the Baobab cluster at University of Geneva and the CHPC. YF was supported by the Robert Bosch Stiftung through the AIMS-ARETE chair position. 
This research has been conducted using resources provided by the United Kingdom Science and Technology Facilities Council (UK STFC) through the Newton Fund and the South African Radio Astronomy Observatory (SARAO). The financial assistance of the SARAO is hereby acknowledged.

\section*{Data Availability}
The data used in this paper is available upon request.


\bibliographystyle{mnras}
\bibliography{ref}



\appendix

\section{\aoflagger\ Optimisation Results }\label{app:aoflagger}

\begin{figure*}
	\centering
	\includegraphics[scale=0.35]{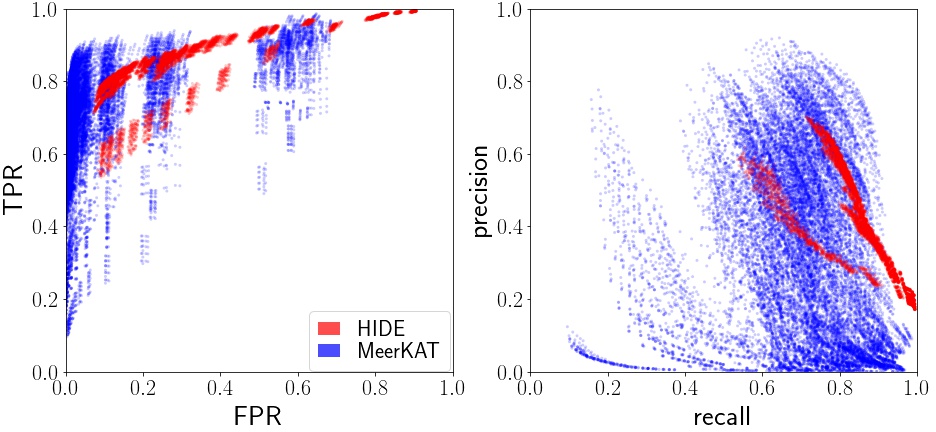}
	\caption{The \aoflagger\ performance in terms of TPR, FPR, recall and precision on both the HIDE (red dots; evaluated on 1 day of data) and MeerKAT simulations (blue dots; evaluated on 800 seconds of data) as the hyper parameters,  $R_b$ (background reject), $\sigma_O$ (outlier n sigma) and $\tau_{n}$ (RFI threshold) are varied widely. There are no combinations that simultaneously lead to both excellent precision, recall and False Positive Rate (FPR).}
	\label{fig:sdp_compare}
\end{figure*}

To provide fair comparison to \rnet\ and \unet\ we explore how the \aoflagger\ results (FPR, recall, precision) are affected by varying the hyperparameters.
We run the \aoflagger\ on HIDE and MeerKAT simulations varying three the hyperparameters ($\tau_{n}$, $\sigma_O$, $R_b$); namely \textit{RFI threshold}, \textit{outlier-n-sigma} and \textit{background reject}, respectively. The first controls  how the user chooses to define the ground truth mask, the second and third are \aoflagger\ hyperparameters. Each single dot in the figures show performance on one simulated file (one day in HIDE simulations and 800 sec for MeerKAT simulations). We show all results as dots in the figures and choose colour to show the dataset. 
Figure \ref{fig:sdp_compare} combines all results in one figure. The results on the MeerKAT simulations are more sensitive to hyperparameter choice.

\section{Metrics}\label{app:metrics}
This appendix gives definitions about the used metrics for who needs to see them in more details.
True Positive Rate (TPR) also called sensitivity or recall is
\begin{equation}
\mbox{Recall} = \frac{TP}{TP+FP}
\end{equation}
where TP (true positive) is the number of pixels that are truly predicted as positive and FP (false positive) is the number of pixels that are mistakenly predicted as positive pixels. The False Positive Rate (FPR), also known as fallout, is 
\begin{equation}
\mbox{FPR} = \frac{FP}{FP+TN}
\end{equation}
where TN (true negative) is the number of the pixels that are truly predicted as negative pixels.  
Then one is able to define the Receiver Operating Characteristic (ROC) curve as TPR plotted against FPR as one changes the classification threshold. The Area Under the Curve (AUC) metric is the integral of the ROC curve. 

The Matthews Correlation Coefficient (MCC) is a number between -1 (worst case) and 1 (perfect case) and is defined by \cite{matthews1975comparison}:
\begin{equation}
\mbox{MCC} = \frac{TP \times TN - FP \times FN}{(TP+FP)(TN+FN)(TP+FN)(TN+FP)}
\end{equation}
where TN and FN are the True Negative and False Negative ratios. It is particularly suited to unbalanced data. 

The Precision is defined by 
\begin{equation}
\mbox{Precision} = \frac{TP}{TP+FP}
\end{equation}
which then allows one to calculate the F1-score using: 
\begin{equation}
F1\mbox{-score} = \frac{2}{\mbox{Precision}^{-1}+\mbox{Recall}^{-1}}
\end{equation}
which mixes the precison and recall metrics. 

\section{Additional Results }\label{app:additional_results}

\begin{figure*}
	\centering
	\includegraphics[scale=0.35]{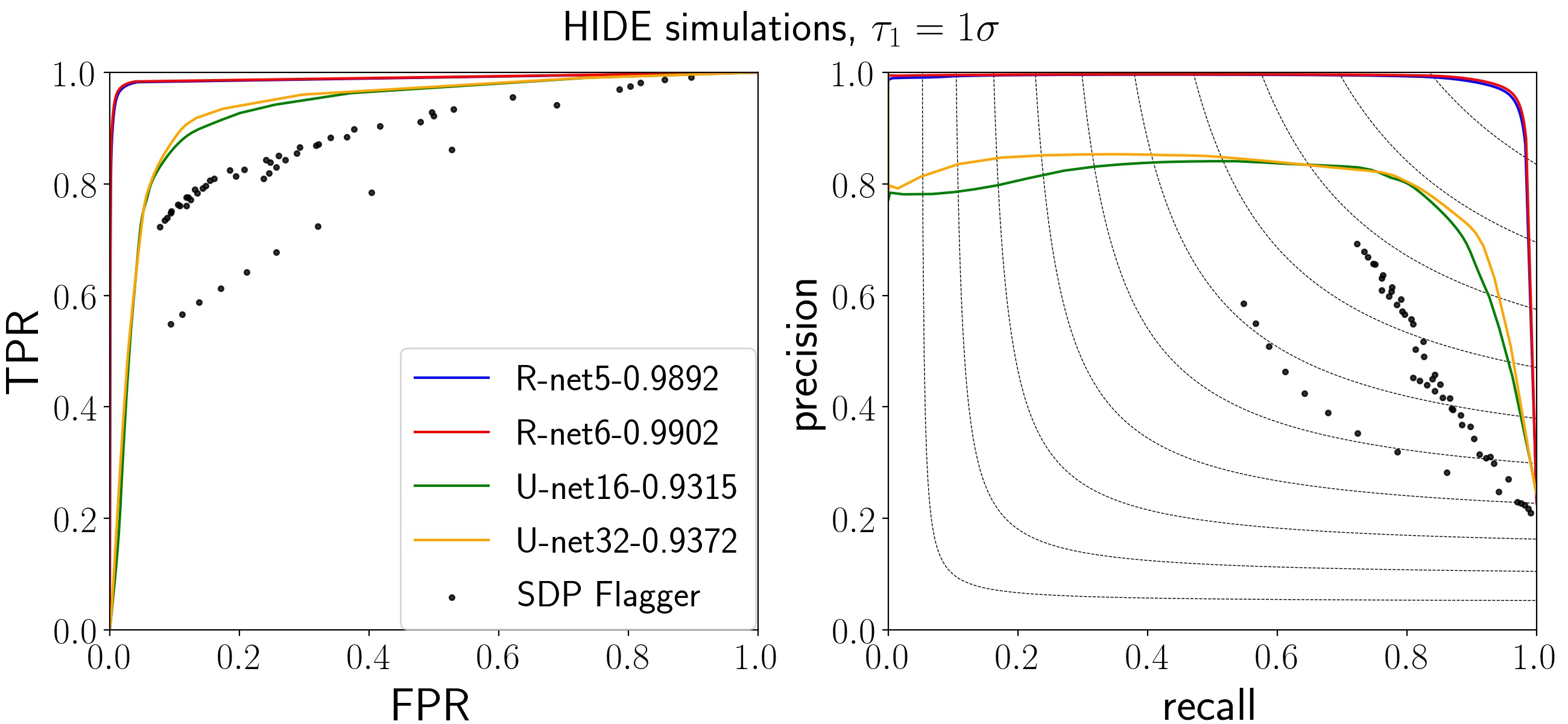}
	\caption{Comparison between \rnet\ and \unet\ performance trained on HIDE simulations to predict for the threshold $\tau_1 = 1\sigma$. Dots show the  \aoflagger\ results  found varying the different hyperparameters. \rnet\ outperforms the other algorithms also for this threshold.}
	\label{fig:hide-cnn2}
\end{figure*}

\begin{figure*}
	\centering
	\includegraphics[scale=0.35]{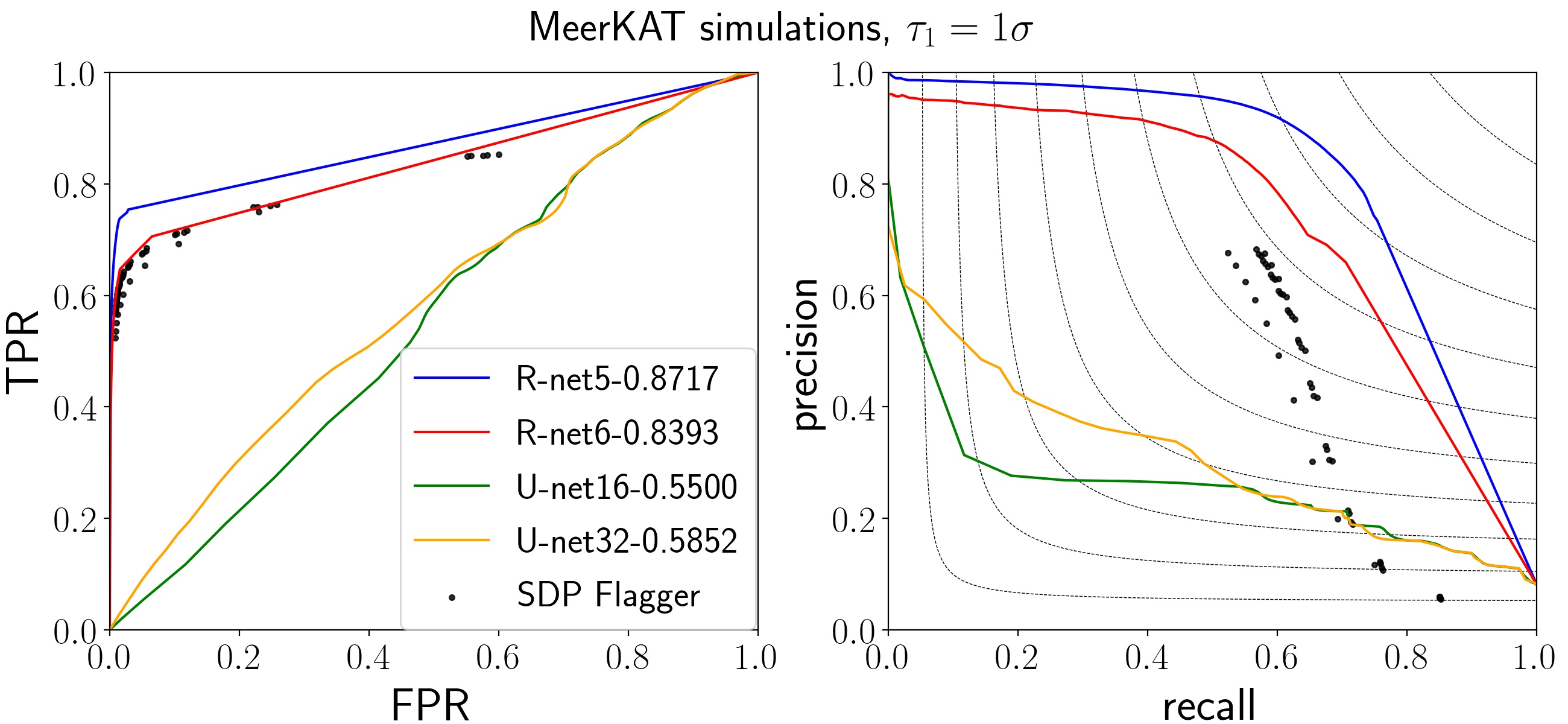}
	\caption{Comparison between \rnet\ and \unet\ performance trained on MeerKAT simulations using only absolute value of the data to predict the threshold $\tau_1 = 1\sigma$. Dots show the \aoflagger\ results using different hyperparameters values. \rnet\ outperforms the other algorithms also for this threshold.}
	\label{fig:mk-cnn2}
\end{figure*}

\begin{figure*}
  	\centering
  	\includegraphics[scale=0.35]{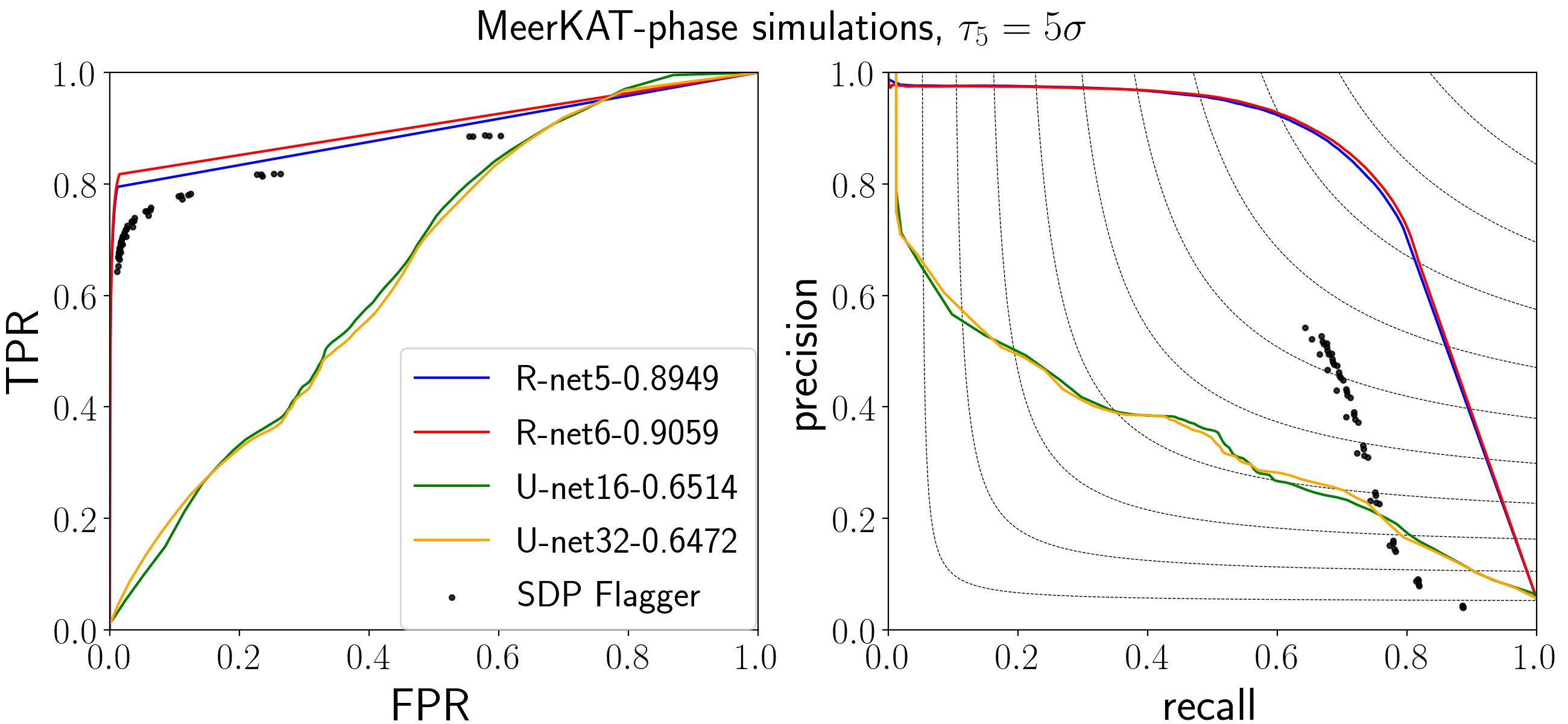}
  	\caption{{\bf Adding phase information yields no improvement - } \rnet\ and \unet\ performance trained on MeerKAT simulations with amplitude and phase of the data to predict threshold=$5\sigma$. Dots are \aoflagger\ results varying $\sigma_O$ and $R_b$. This should be compared with Fig. \ref{fig:mk-cnn1} which shows almost identical performance for \rnet, though modest improvement for \unet.}
  	\label{fig:mkph-cnn1}
  \end{figure*}

In this appendix we show the results for the HIDE and MeerKAT data for a threshold of $\tau_1 \equiv 1\sigma$ in Fig. \ref{fig:hide-cnn2} and \ref{fig:mk-cnn2}. In addition, in Fig. \ref{fig:mkph-cnn1} we show the effect of adding uncalibrated phase information as a new channel to the visibility magnitude maps on the performance of \rnet\ and \unet. Comparison with Fig. \ref{fig:mk-cnn1} shows that very little is gained in terms of AUC for \rnet. 

\begin{figure}
	\centering
	\includegraphics[scale=0.38]{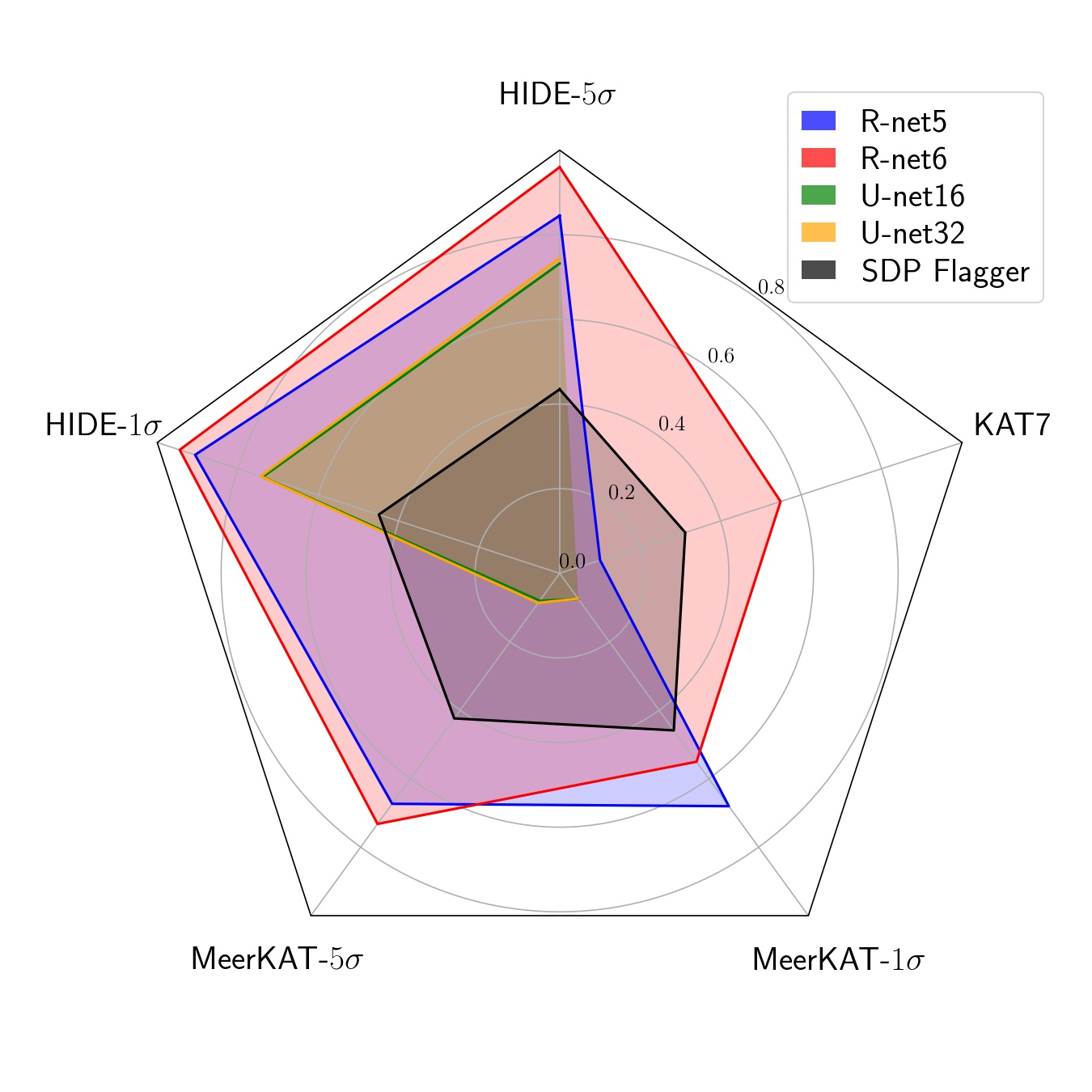}
	\caption{Starplot comparison between the MCC scores of all the algorithms for all the datasets and different thresholds (e.g. HIDE-$5\sigma$ corresponds to the HIDE results where the threshold is $\tau_5$.)  }
	\label{fig:starplot_mcc}
\end{figure}

\begin{figure}
	\centering
	\includegraphics[scale=0.38]{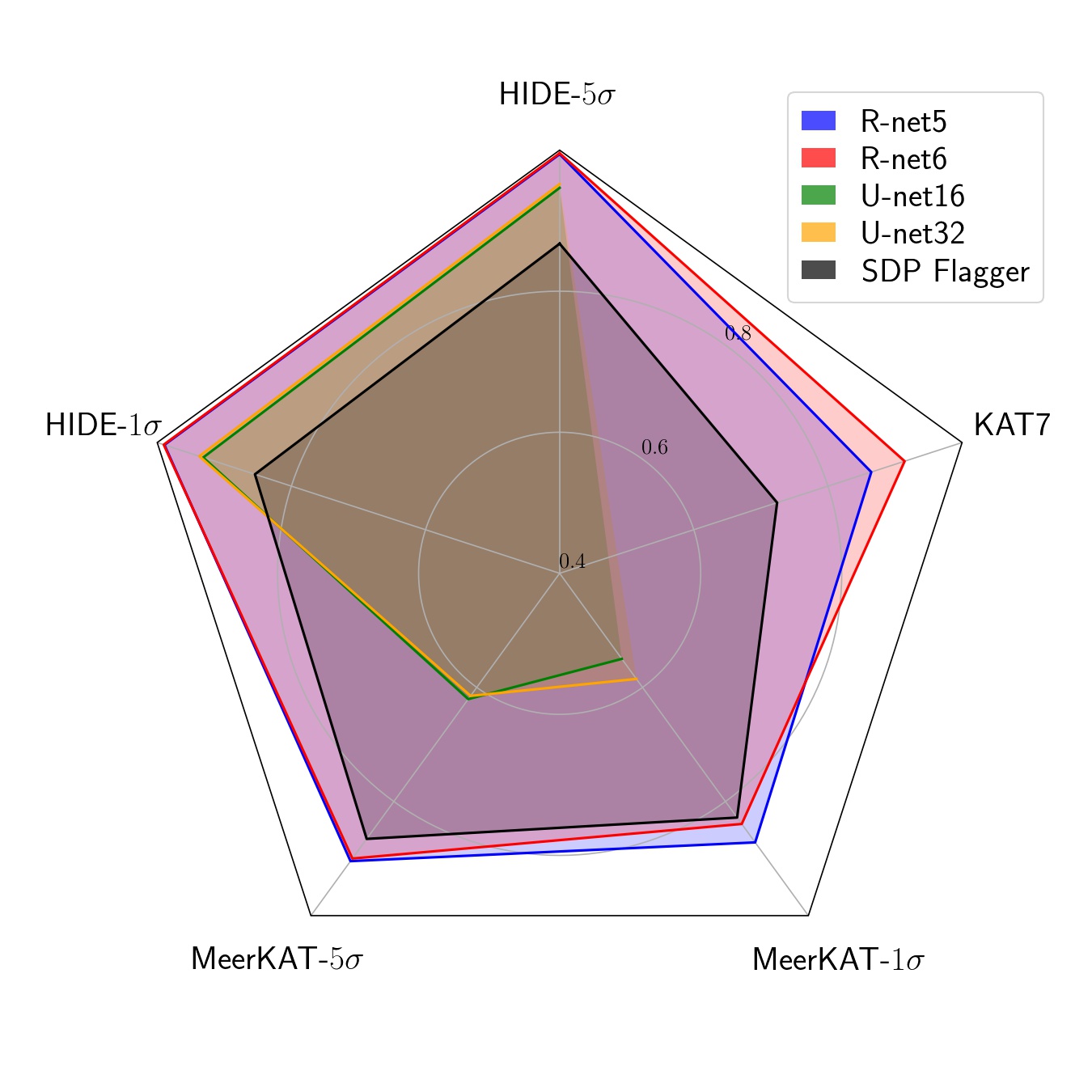}
	\caption{Starplot comparison between AUC scores of all the algorithms for all the datasets and different thresholds (e.g. HIDE-$5\sigma$ corresponds to the results where the threshold is $\tau_5$).  \rnet\ outperforms all other algorithm variants. The \aoflagger\ AUC is estimated by interpolating points in the TPR-FPR plane. 
	}
	\label{fig:starplot_auc}
\end{figure}

Finally, in the starplots \ref{fig:starplot_mcc} and \ref{fig:starplot_auc} we compare the performance of all the algorithms on all the datasets for both combinations of threshold, $\tau_1$ and $\tau_5$. 

\bsp	
\label{lastpage}
\end{document}